\documentclass[pre,epsfig,aps,twocolumn]{revtex4}

\usepackage{epsfig}
\usepackage{fancyheadings}
\usepackage{amstext,amsmath,amssymb}

\begin{document}

\sloppy
\pagestyle{empty}

\title{Reconstructing a Random Potential from its Random Walks}
\author{S. Cocco $^1$, R. Monasson $^2$}
\affiliation{$^1$ 
CNRS-Laboratoire de Physique Statistique de l'ENS, 24 rue Lhomond,
75005 Paris, France\\
$^2$ CNRS-Laboratoire de Physique Th\'eorique de l'ENS, 24 rue Lhomond,
75005 Paris, France}

\begin{abstract}
The problem of how many trajectories of a random walker in a potential
are needed to reconstruct the values of this potential is
studied. We show that this problem can be solved by calculating 
the probability of survival of an abstract random walker in a
partially absorbing potential. The approach is illustrated on the 
discrete Sinai (random force) model with a drift.
We determine the parameter (temperature, duration of
each trajectory, ...) values making reconstruction as fast as
possible. 
\end{abstract}

\maketitle

{\em Introduction.}
Random walks (RW) in random media have been intensively studied in the
past decades as a paradigm for out-of-equilibrium dynamics, and have 
led to the discovery and understanding of
important dynamical effects as anomalous diffusion, ageing
...\cite{discreteRF,revue}. Briefly speaking the issue is to determine the
statistical properties of the walker from the ones of the energy
potential. Much less attention has been devoted to the inverse
problem: given one (or more) observed RW(s) can we guess the potential
values?  
This question naturally arises in biophysics where the use of
 AFM, optical and magnetic tweezers make  possible the
mechanical separation of single protein-protein complexes \cite{evans}, 
or the unfolding and refolding of single biomolecules\cite{fer04,Ess97,woo06}.
The observed dynamics the rupture of chemical bonds, 
of folding/unfolding of nucleic acids,
or proteins   can be modeled as a RW motion affected by thermal
noise, moving in a quenched potential determined by  
the composition of the chemical bonds, or the sequence of
amino-- or nucleic--acids. Reconstructing the free
energy landscape of those processes is the object of current and
intense efforts  \cite{hye03,evans,mb,ritort,woo06}. 

In this letter we show how the inverse RW problem
can be practically solved within the Bayesian inference framework and
address the crucial question of the accuracy of reconstruction.
In practice information can be accumulated either by increasing the 
duration of one RW, or observing more than one RW, or combining the two.
We discuss the optimal procedure minimizing the total number of
data to be acquired, and show how this minimal amount of data can be
calculated from
the probability of survival of an abstract  walker in a partially 
absorbing potential. The approach is illustrated in detail on the celebrated 
discrete random force (RF) model (Sinai model with non zero drift)
\cite{revue,discreteRF}. 

Inference is a key issue in information theory and
statistics \cite{bayes}, with applications in biology \cite{domany}, 
social science \cite{And57}, finance,  ... A central question is the
so-called hypothesis testing problem: which one of two candidate 
distributions is
likely to have generated a set of measured data?  This question was 
solved in the case of independent variables by Chernoff \cite{Che52}, 
and is the core issue of the asymptotic theory of inference
\cite{bayes}. Chernoff showed that the probability of guessing 
the wrong distribution decreases exponentially with the size of the data
set \cite{Che52}. Large deviations techniques can be used to treat 
the case of variables extracted from one recurrent 
realization of a finite Markov chain \cite{Boz71,Dem98}; the present
work can be seen as an extension to many transient realizations of
an `infinite' chain.

\begin{figure}
\begin{center}
\vskip .7cm
\psfig{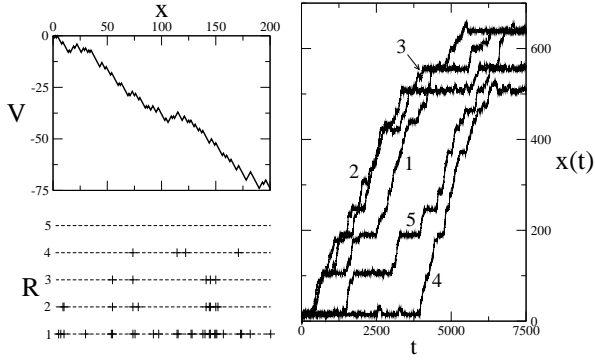}
\vskip .7cm
\caption{Left, top:  Example of  potential ${\bf V}$ obtained in the RF model
with tilt $b=0.4$ (size $N=1000$, 
sites $x>200$ not shown here).  Right: examples of RWs, numbered 
from 1 to 5, in this potential at temperature $T=1$; plateaus are in
correspondence with the local minima of $V$. Here $\alpha \simeq 0.85$
(creep phase). Left, bottom: Predictions from the first $R$ RWs in the
right panel and (\ref{field}); 
impulses locate incorrectly predicted forces $f_x$ for $x\le
200$. The number of erroneous forces decreases from 26 (for $R=1$) to
0 ($R=5)$. Note the errors on sites $x_0\simeq
100$ appearing when the fourth RW is taken into account; indeed this 
atypical RW marks no pause in the local minimum in $x_0$. }
\label{fig-pot}
\end{center}
\end{figure}

{\em Random Force model.} For an illustration of the problem
consider the discrete, one dimensional RF model defined on the set
of sites $x=0,1,2,\ldots ,N$ \cite{discreteRF}. We start by choosing
randomly a set of dimensionless forces $f_{x}=\pm 1$ on each link
($x,x+1$) with {\em a priori}
probability $P_0 =\prod_x \frac{1+ b\,f_x}2$ where $-1<b<1$ is
called tilt. This defines the values of the potential ${\bf V}$ on each site,
$V_x = - \sum_{y<x} f_y$ (by definition $V_0=0$). 
An example of potential for $b=0.4$ is shown on Fig.~\ref{fig-pot}.

After the quenched potential has been drawn
a random walker starts in $x=0$ at time $t=0$. The walker then 
jumps from one site $x$ to one of its neighbors $x'=x\pm 1$ with
rate (probability per unit of time) $r_{\bf V}(x \to
x')=r_0\times e^{(V_x-V_{x'})/(2T)}$ to satisfy detailed balance
at temperature $T$; the attempt rate $r_0$ will be set to unity in the
following. Reflecting boundary conditions are imposed by
setting $V_{N+1}=V_{-1}=+\infty$. We register the sequence of 
of positions up to some time $t_f$: ${\bf X}=\{ x(t), 0\le
t\le t_f\}$. Figure~\ref{fig-pot} shows five RWs ${\bf X}_\rho$,
$\rho=1,\ldots,5$ , each starting in the origin $x(0)=0$ and of equal 
duration $t_f$ for a temperature $T=1$. The value of the temperature
strongly affects the dynamics \cite{revue}, and its relevance for
the inverse problem will be discussed later. 

Our objective is to reconstruct the potential over a region of the
lattice e.g. the value of the forces on some specific links
from the observation of RWs. 
Within Bayes inference framework this can be done by maximizing
 the joint probability of the potential ${\bf V}$ 
and of the observed RWs ${\bf X}_1,\ldots, {\bf X}_R$
 over ${\bf V}$ \cite{bayes}. 
$P$ is the product of the {\em a priori} probability of the potential,
$P_0$, times the likelihood of
the RWs given the potential, $L$. Since the RW is Markovian $L$ depends
only on the sets of total times $t_x$ spent on every site $x$, 
and of the numbers of jumps $u (x\to x')$ from
$x$ to $x'$ over the set of RWs: 
\begin{equation} \label{like}
L =  \prod_{x,x'} e^{-t_x \;r_{\bf V}(x\to x')}
\;r_{\bf V}(x\to x') ^{u(x\to x')}
\end{equation}
where the product runs over all sites $x$ and their neighbors $x'=x\pm 1$.
Expressing the rates in terms of the forces and maximizing the joint
probability $P$ we obtain  the most likely values for the forces:
$f_x=\mbox{sign}(h_x + \alpha)$ where $\alpha\equiv T\, \ln[
  (1+b)/(1-b)]$ is a global `field'
coming from the {\em a priori} distribution $P_0$ and $h_x$ a local 
contribution due to the likelihood $L$, 
\begin{equation} \label{field}
h_x = 2T\sinh \big(\frac 1{2T}\big) \; ( t_{x+1} -t_x)  + u (x\to x+1)
 - u (x+1\to x) \ .
\end{equation}
Figure~\ref{fig-pot} (left, bottom) shows predictions made from
$R=1$ to $R=5$  RWs for the first 200 sites. 
The duration $t_f$ of the RW is chosen 
to be much larger than the mean first passage time
in $x=200$, and much smaller than the equilibration time 
$t_{eq}\sim e^{bN/T}$. In this range the quality of prediction is essentially 
independent of $t_f$ as will be discussed in detail below. 
As expected the number of erroneous forces 
decreases with increasing $R$ though atypical events may
produce flaws in the prediction. The analysis of these atypical RWs, 
and how they lead to errors is the keystone of what follows.

{\em Number of RWs necessary for a good reconstruction.} 
Expression (\ref{like}) for the likelihood of the RWs 
is true for any potential
${\bf V}$ and can be geometrically interpreted as follows.
Given a set of RWs we extract a signal vector ${\bf S}$ 
whose components are:  the times $t_x$ spent on
site $x$, the numbers $u(x\to x')$ of transitions 
from site $x$ to site $x'$. When $R$ is large we expect ${\bf
S}$ to be extensive with $R$ and define the intensive signal
${\bf s}={\bf S}/R$. 
Similarly, to each potential ${\bf V}$ we associate a vector
${\bf v}$ with components: minus the outgoing rate {\em i.e.} $ -\sum
_{x' (\ne x)} r_{\bf V} ( x\to x')$ for each site $x$, the logarithm of
the rate $r_{\bf V}(x\to x')$ for each pair of neighbors. 
Then $L=\exp(R \;{\bf s}\cdot {\bf v})$ from (\ref{like}) 
where $\cdot$ denotes the scalar product. Maximizing the joint
probability $P=P_0\times L$ over the potential becomes equivalent,
in the large $R$ limit, to finding ${\bf v}$ with the largest
scalar product with the signal ${\bf s}$
\footnote{The irrelevance of the {\em a priori} distribution 
in the asymptotic case of large data set is well-known \cite{bayes}
and can be checked for the RF model: the local field (\ref{field})
is extensive in $R$, while the global field $\alpha$ remains 
finite.}.
It is natural to partition the space of signals into `Voronoi cells': 
$C_{\bf v}$ is the set of ${\bf s}$ having a larger scalar product with
${\bf v}$ than with any other potential ${\bf v}'$. 
Bayes rule tells us that the most likely potential given an observed
signal ${\bf s}$ is the one attached to the cell in which ${\bf s}$ lies.

Consider now RWs taking place in a given potential ${\bf V}$.
From the law of large number the signal ${\bf s}$ is equal, in 
the infinite $R$ limit, to 
${\bf s}^*_{\bf v}=\{t_x^*,u^*(x\to x')=t^*_x \, r_{\bf V} (x\to x')\}$ where
$t^*_x$ is the average sojourn time on site $x$ 
over RWs of duration $t_f$. As ${\bf s}^*_{\bf v} \in C_{\bf
  v}$ \footnote{Let ${\bf v}'\ne {\bf v}$;
${\bf s}^*_{\bf v}\cdot ({\bf v}-{\bf v}') = \sum _{x\ne x'} 
u^* (x\to x') G( r_{{\bf V}'}(x\to x')/ r_{{\bf V}}(x\to x') )$ where
$G(z)=z-\ln z -1 >0$ for $z\ne 1$.
} reconstruction becomes flawless in the limit
of an infinite number of data  as expected. 
For large albeit finite $R$, ${\bf s}$ typically deviates
from ${\bf  s}^*_{\bf v}$ by $O(R^{-\frac 12})$; finite deviations have
exponentially small--in--$R$ probabilities, $e^{-R \, \omega _{\bf V}
  ({\bf s})}$, controlled by a 
rate function $\omega _{\bf V} ({\bf s})$ \cite{Dem98}. 
The probability to predict an erroneous potential is 
the probability that the stochastic signal ${\bf s}$ does not 
belongs to cell $C_{{\bf v}}$. This probability of error thus decays 
exponentially with $R$ over a typical number of RWs
\begin{equation} \label{min}
R_c({\bf V}) = \big[\ \displaystyle{\min _{{\bf s} \notin C_{\bf v}}} 
\ \omega _{\bf  V} ({\bf s})\ \big]^{-1}  \ ,
\end{equation}
where the minimum is taken over signals outside the `true' 
cell. It depends on the temperature, the duration of the RW, ...

As the RWs are independently drawn $\omega _{\bf   V}$ is a convex
function of ${\bf s}$ \cite{Dem98}. The minimum in (\ref{min}) is thus
reached on the boundary between the true cell and
another, bad cell, say, $C_{\bar {\bf v}}$. The attached potential, 
$\bar {\bf V}$, is the most `dangerous' one from the inference point of
view. RWs generated from ${\bf V}$ and $\bar {\bf V}$ are hardly told
from each other unless more than $R_c({\bf V})$ of them are observed. 

Assume $\bar {\bf V}$ is known. Then the boundary between $C_{\bf v}$
and $C_{\bar {\bf v}}$ is the set of signals ${\bf s}\perp {\bf v}
-\bar {\bf v}$. We deduce
\begin{equation}
\label{bound}
R_c({\bf V} ) = \big[\ \max _{\mu} \; \min _{{\bf s}} 
\big( \omega _{\bf V} ({\bf s}) + \mu\, {\bf s}\cdot( 
 {\bf \bar v}-{\bf  v}) \big)\ \big] ^{-1}\ 
\end{equation}
where the Lagrange multiplier $\mu\in [0;1]$ ensures that ${\bf s}$ is confined
to the boundary.
The Legendre transform of $\omega _{\bf V}$ appearing in (\ref{bound})
is intimately related to the evolution operator of an abstract
random  walk process, denoted by RW$(\mu)$
to distinguish from the original RW \cite{noi}. This RW$(\mu)$-er 
moves with the rates 
$r_{(1-\mu) {\bf V} + \mu \bar {\bf  V}} (x \to x')$ and may die
on every site $x$ with positive rate 
\begin{eqnarray} \label{death}
&&d_{{\bf V},\bar {\bf V},\mu} (x) 
=\sum _{x' (\ne x)}  \big[ (1-\mu)\, r _{\bf V} (x\to x') +
\mu \,  r_{\bf \bar V}(x\to x') \nonumber \\
 &&\hskip 3cm  -\ r_{(1-\mu) {\bf V} + \mu {\bf \bar V}}(x\to x')\big] \ .
\end{eqnarray}
Consider now the probability $\pi(\mu)$ 
that RW$(\mu)$-er, initially at the origin,
has survived up to time $t_f$ (the duration of the original RW).
Then
$R_c({\bf V}) =\displaystyle{\min _{\mu \in[0;1]} 1/| \ln \pi(\mu)|}$. 

{\em Optimal Working Point for the RF model.} 
We apply the above theory to the discrete RF model, and want to predict 
the value of the force $f_y$ on the link $(y,y+1)$ for some specific
$y$. The dangerous potential is ${\bf \bar V}$ 
obtained from ${\bf V}$ upon reversal of the force $f_y\to -f_y$.
We aim at calculating the probability $\pi(\mu)$ of survival of 
RW$(\mu)$-er moving with rate $r(x\to x')=r_{\bf V} (x\to x')$
and dying on site $x$
with rate $d(x)=0$ except:  $r(y\to y+1)
=1/r(y+1\to y)=e^{(1-2\mu)f_y/(2T)}$,
$d(y)= D(f_y),  d(y+1)=D(-f_y)$ where
$D(f)\equiv (1-\mu) e^{f/(2T)}+\mu e^{-f/(2T)} - e^{
(1-2\mu) f/(2T)}$ from (\ref{death}). From the
previous section the number of RWs required for a reliable prediction
of $f_y$ is $R_c(y;{\bf V}) = \min _{\mu} 1/|\ln \pi(\mu)|$.

\begin{figure}
\begin{center}
\vskip .7cm
\psfig{figure=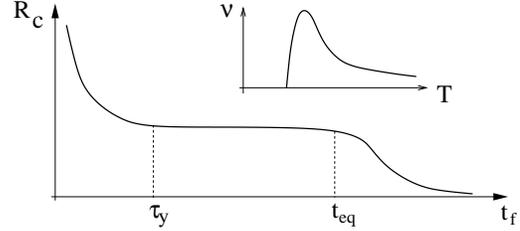,height=3cm,angle=0}
\caption{Sketch of the number $R_c(y;{\bf V})$ of RWs 
necessary for a good inference of the force $f_y$ 
as a function of the RW duration $t_f$. $\tau_y$ is the typical 
first-passage time in $y$ from the origin, $t_{eq}$ the equilibration
time  (comparable to the first-passage time from the extremity
$N$ when $y\ll N$). Inset: rate of reconstruction (\ref{vel}) as
a function of temperature at fixed tilt.} 
\label{fig-rc}
\end{center}
\end{figure}

Let $\pi _x(\mu,t)$ be the
probability that RW$(\mu)$, initially on site $x$,  is still 
alive at time $t$. The time-evolution of $\pi_x$ is described by
\begin{equation} \label{diffeq}
\frac{\partial\pi _x }{\partial t} = \sum
_{x' (\ne x)}  r(x\to x') \big( \pi _{x'}- \pi_{x} \big) 
-d (x)\, \pi_x \ ,
\end{equation}
with initial  condition $\pi_y (\mu,0)=1$ (by convention
$\pi_{-1}=\pi_{N+1}=0$). After Laplace transform over time, eqns 
(\ref{diffeq}) are turned into recurrence equations for the ratios
$\pi_x/\pi_{x+1}$ and solved with great numerical accuracy. We obtain this way
the probability of survival, $\pi(\mu)=\pi_0(\mu,t_f)$, and optimize
over $\mu$.
Though $R_c$ depends on the potential ${\bf V}$ its general behavior
for tilt $b>0$ as a function of the duration $t_f$ is sketched in
Fig.~\ref{fig-rc}. Three regimes are observed: 

$\bullet$ for $t_f\ll \tau_y$ (mean first passage time in $y$)  
RW$(\mu)$ has a low probability to visit $y$ and is almost surely alive,
hence $R_c$ is very large; 

$\bullet$ for $\tau _y\ll t_f\ll t_{eq}$ RW$(\mu)$ has visited 
the region
surrounding $y$ and escaped from this region (transient regime), hence
its probability of survival remains constant, and so does 
$R_c$; 

$\bullet$ for $t_f\gg t_{eq}$ RW$(\mu)$ visits again
and again the region surrounding $y$, hence 
the probability of survival decreases exponentially
with the duration: $R_c \propto 1/t_f$. 

The total time $R_c\times t_f$ for a good reconstruction 
is minimal when we choose $t_f \gtrsim \tau_y$. This
marginally transient regime corresponds to the plateau of
Fig.~\ref{fig-rc}: RWs are long enough to 
visit site $y$ but short enough not to wander much away from $y$. 
To calculate the corresponding value of $R_c$ we take 
the limits, in order, $N\to\infty$, $t_f\to \infty$, and look for the
stationary solution of (\ref{diffeq}) with boundary condition
$\pi_{x\to\infty}=1$. The result for
the probability of survival is
\begin{equation} \label{pstarsinai}
\pi (\mu) = \frac{e^{-\frac{\mu}T}}{1-\mu+ \mu 
\, e^{-\frac 1T}+\mu(1-\mu)\, t^* _{y+1}\,(e^{\frac{1}{4T}}- 
e^{-\frac{3}{4T}})^2 } \ ,
\end{equation}
where the mean sojourn time on site $y+1$ in ${\bf V}$ is \cite{revue}
\begin{equation} \label{tx}
t ^*_{y+1}= \sum _{z\ge0} \exp \left[\frac 1{T} \left( \frac { 
V_{y+z+2}+ V_{y+z+1}}2 - V_{y+1} \right)\right] \ .
\end{equation}

\begin{figure}
\begin{center}
\vskip .7cm
\psfig{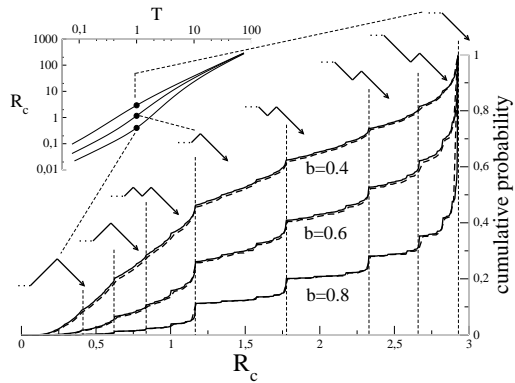}
\caption{Cumulative probability distribution ${\cal Q}$ of 
$R_c(y;{\bf V})$ at
temperature $T=1$ and for three tilt values $b$. Full lines are
numerical results from $10^6$ samples, and dashed lines are the
outcomes from the $\ell$-pattern approximation. Inset: $R_c$ vs. 
$T$ for the 3-patterns  $+++$,
$-++$, $---$ (from top to down).}
\label{fig-histo}
\end{center}
\end{figure}

{\em Distribution of $R_c$ over potentials.}
The number $R_c(y;{\bf V})$ of RWs necessary to predict the value of
$f_y$ depends on the potential ${\bf V}$ through the sojourn time 
$t^*_{y+1}$ (\ref{tx}). By randomly drawing potentials (or varying site $y$) 
we obtain the distribution of $R_c$ shown in Fig.~\ref{fig-histo}. 
Main features are:

$\bullet$ Small $R_c$ correspond to sites where the RW spends long
time $t^*$ (traps)\footnote{RW$(\mu)$, due to conditioning to survival, 
is likely to stay for $\sim 1/d(y) \ll t^*$ in the trap only.}: 
$R_c \sim \frac 1{|\ln \pi|} \sim \frac 1{\ln t^*}$ from 
(\ref{pstarsinai}). The power law tail of the distribution of 
sojourn times, $P(t^* )\sim (t^*) ^{-(\alpha+1)}$ \cite{revue}, 
gives rise to an essential singularity at the origin 
in the cumulative distribution, ${\cal Q} (R_c) \sim
e^{-\alpha/R_c}$. The potential is easy to predict over trapping
regions since RWer spends a long time there, and accumulates
information about the energy landscape.

$\bullet$ Conversely the largest value of $R_c$, denoted by $R_c ^H$,
correspond to the homogeneous potential $V_x^H=-x$ in which the
walker is never  trapped and is quickly driven to $+\infty$.
$R_c^H$ can be calculated from (\ref{pstarsinai}) by setting
$f_x=+1$ for all sites in (\ref{tx}). The singularity in 
${\cal Q}$ when $R_c\to R_c^H$ corresponds to quasi-homogeneous
potentials, where one force, say, on site $\ell$, is $-1$.
Such potentials have exponential-in-$\ell$ small probabilities, but
give values of $R_c$ on site $y=0$ exponentially close to $R_c^H$. 
On the overall we
find $1-{\cal Q} (R_c ^H - \epsilon) \sim \epsilon ^\beta$ where
the exponent is $\beta=T \ln \frac {1+b}2$. 

$\bullet$ In between ${\cal Q}$ shows marked steps at well defined and
$b$-independent values of $R_c$, which correspond to specific 
local force patterns beyond site $y$.  
A $\ell$-pattern is defined as a sequence of
forces on sites $y+1$ to $y+\ell+1$, followed by all $+$
forces; the corresponding  $R_c$ can be
exactly calculated from (\ref{pstarsinai},\ref{tx}), and is 
shown  for 7 among the
16 $\ell=4$-patterns in Fig.~\ref{fig-histo}. 
The histogram of $R_c$ can be accurately approximated for any tilt $b>0$ based
on the above local pattern description. Given a length $\ell$ we enumerate all
the $2^\ell$ patterns, calculate the corresponding $R_c$, and weight them
with probability $(\frac {1+b}2) ^{\# f_x =+}\times (\frac {1-b}2) 
^{\# f_x =-}$. In practice we choose $\ell\sim 10/\ln [2/(1-b)]$,
to ensure that patterns with more than $\ell$ negative 
forces have negligible
weights ($< e^{-10}$). The resulting histograms are in 
excellent agreement with ${\cal Q}$ for intermediate
values of $R_c$ (dashed lines in  Fig.~\ref{fig-histo}). 

{\em Tuning temperature for fast reconstruction.}
The dependence of $R_c$ upon temperature is shown for three patterns 
in the Inset of Fig.~\ref{fig-histo}. We have $R_c \sim 4T$
as $T\to\infty$ independently of the pattern, and $R_c\sim 2T/(h+3)$
when $T\to 0$ where $h$ is the highest barrier to the right of $y$
in the potential defined by the pattern (Fig.~\ref{fig-histo}). 
When the temperature exceeds the temperature $T_b$ such that 
$\alpha=1$ the velocity 
of the RWer is finite $\frac y{\tau _y} \sim v(T) >0$ 
\cite{revue}. The reconstruction rate (number of
correctly predicted forces per unit of time) is equal to the velocity
$v(T)$ divided by $R_c$, 
\begin{equation} \label{vel}
\nu (T)= \frac{1-\cosh\frac 1T + b \sinh
\frac 1T}{\cosh \frac 1{2T} - b \sinh \frac 1{2T}} \times \int _0 ^{R_c^H} 
dR_c \frac {{\cal Q}  '(R_c)}{R_c} 
\end{equation}
after averaging over the quenched potential.
The dependence of $\nu$
upon temperature is sketched in the Inset of Fig.~\ref{fig-rc}; it is
maximal and equal to $\nu ^M$ for some temperature $T^M$ realizing a
trade-off between fast motion (large velocity) and accurate
reading-out (small $R_c$). Even in the small
$b$ limit the optimal reconstruction rate is finite, $\nu^M \sim b^2$, 
by working at high temperature $T^M \sim \frac  1b$, while in the
absence of optimization procedure the number of predicted forces
scales only as the squared logarithm of the time \cite{math2}.

{\em Conclusion.} We have shown how the number of RWs
required for a good reconstruction of the potential
can be deduced from the probability of
survival of an absorbing RW process. This result is of practical
interest since the survival probability can be estimated through 
numerical simulations e.g. in dimension $\ge 2$. Furthermore we have
determined, for the special case of the RF model, the optimal
`experimental' protocol for reconstruction (number of RWs,  duration, 
temperature).

Our formalism applies to continuously parametrized potentials 
e.g. RF model with forces taking continuous
instead of binary values. The aim is now to predict the true potential 
values up to some accuracy on each site; this is turn 
determines an acceptable neighborhood around ${\bf s}^*_{\bf v}$ in the space
of signals. The rate function $\omega _{\bf v}$ is generically
parabolic around ${\bf s}^*_{\bf v}$, with a curvature matrix called Fisher
information matrix \cite{bayes}. Finding $R_c$ amounts to minimize this
(positive) quadratic form on the boundary of the neighborhood, a task 
which can be carried out efficiently \cite{garey}. 
Our approach can be easily  extended to the
case of a finite delay between two measures of the positions, and
Chernoff's result is recovered in the finite $N$, infinite delay 
limits \cite{Che52,mb}. 

{\em Acknowledgments.} We are grateful to D. Thirumalai for his 
suggestion of illustrating our formalism on the RF model. This work 
was partially funded by ANR under  contract 06-JCJC-051.

\end{document}